\documentclass[twocolumn]{aastex63}
\usepackage{ amsmath}

\newcommand{\source}{FRB 200428}

\submitjournal{ApJL}

\shorttitle{Neutrino Counterparts of FRBs}
\shortauthors{Metzger, Fang, Margalit}
\graphicspath{{./}{figures/}}

\begin{document}

\title{\Large{Neutrino Counterparts of Fast Radio Bursts}}

\correspondingauthor{Brian D.~Metzger}
\email{bdm2129@columbia.edu}

\author{Brian D.~Metzger}
\affiliation{Department of Physics, Columbia University, New York, NY 10027}
\author{Ke Fang} 
\altaffiliation{NHFP Einstein Fellow}
\affil{Kavli Institute for Particle Astrophysics and Cosmology (KIPAC), Stanford University, Stanford, CA 94305, USA}

\author{Ben Margalit}
\altaffiliation{NHFP Einstein Fellow}
\affiliation{Astronomy Department and Theoretical Astrophysics Center, University of California, Berkeley, Berkeley, CA 94720, USA}

\begin{abstract}

The discovery of a luminous radio burst, \source,  with properties similar to those of fast radio bursts (FRB), in coincidence with an X-ray flare from the Galactic magnetar SGR 1935+2154, supports magnetar models for cosmological FRBs.  The burst's X-ray to radio fluence ratio, as well as the X-ray spectral shape and peak energy, are consistent with \source~being the result of an ultra-relativistic shock (powered, e.g., by an ejected plasmoid) propagating into a magnetized baryon-rich external medium; the shock simultaneously generates X-ray/gamma-rays via thermal synchrotron emission from electrons heated behind the shock, and coherent radio emission via the synchrotron maser mechanism.  Here, we point out that a unique consequence of this baryon-loaded shock scenario is the generation of a coincident burst of high-energy neutrinos, generated by photo-hadronic interaction of relativistic ions$-$heated or accelerated at the shock$-$with thermal synchrotron photons.  We estimate the properties of these neutrino burst FRB counterparts and find that a fraction $\sim 10^{-8}-10^{-5}$ of the flare energy (or $\sim 10^{-4}-10^{-1}$ of the radio isotropic energy) is channeled into production of neutrinos with typical energies $\sim$ TeV$-$PeV.  We conclude by discussing prospects for detecting this signal with IceCube and future high-energy neutrino detectors.
\end{abstract}

\keywords{
Radio transient sources (2008) --- Magnetars (992) --- Soft gamma-ray repeaters (1471)
}

\section{Introduction} 
\label{sec:intro}

Fast radio bursts (FRBs) are millisecond duration pulses of coherent radio emission of extragalactic origin (\citealt{Lorimer+07,Thornton+13}; see \citealt{Petroff+19}, \citealt{Cordes&Chatterjee19} for reviews).  Although many models have been proposed for FRBs (\citealt{Platts+19}), their central engines and emission mechanism remain elusive.  The most well-studied class of models, which can for instance accommodate the discovery that many FRBs are observed to recur on timescales from minutes to weeks (e.g.~\citealt{Spitler+16,CHIME+19}), postulate an origin associated with flaring activity of magnetars \citep{Popov&Postnov13,Lyubarsky14,Kulkarni+14,Katz16,Murase+16,Metzger+17,Beloborodov17,Kumar+17}.  

Magnetar FRB models received a recent  jolt of support following the discovery of a double-peaked radio burst in spatial and temporal coincidence with a double-peaked hard X-ray outburst from the Galactic magnetar SGR 1935+2154 \citep{Bochenek+20,CHIME+20}.  The energy radiated by the burst (hereafter FRB 200428) in the $\sim$ GHz radio band, $\mathcal{E}_{\rm radio} \sim 10^{34}-10^{35}$ erg, exceeds by several orders of magnitude that of any radio burst detected prior from a Galactic neutron star (including giant pulses from young pulsars like the Crab) and is within a factor of $\lesssim 10$ of the least energetic cosmological FRBs of known distance.  

The rate of bursts with similar energy and occurrence rate to \source~is itself insufficient to explain the total extragalactic FRB population, as inferred from the measured all-sky FRB rate.  Nevertheless, if the same physical processes can be scaled-up to more energetic bursts, allowing for a more extreme range of magnetar activity than seen in the Milky Way (as needed to explain particularly active recurrent sources like FRB 121102; \citealt{Spitler+16}), such a unified scenario can account for most or all of the cosmological FRB population \citep{Margalit+20,Lu&Piro20}.  

Despite the breakthrough discovery of \source, several open questions remain, particularly on the theory side.  Chief among these is how to distinguish between several proposed variations of the magnetar model, which make qualitatively different predictions for the mechanism and location of the radio emission and the expectation (if any) of each for an accompanying higher-frequency afterglow.  Unlike cosmological FRBs, for which no unambiguous multi-wavelength counterpart has yet been detected, the joint detection of X-ray and radio emission from \source~find a natural explanation in some magnetar models, while placing others in tension \citep{Margalit+20b,Lu+20}.  

Arguably the most well-developed of these models, in terms of making quantitative predictions for the burst properties motivated by first-principles kinetic plasma simulations, postulate that FRBs are the electromagnetic wave precursors of transient magnetized shocks generated by flare ejecta (e.g.~an ultra-relativistically expanding plasmoid) released from the magnetosphere \citep{Lyubarsky14,Beloborodov17}.  In these scenarios the radio precursor is generated at the shock front by electrons undergoing the synchrotron maser process (e.g.~\citealt{Gallant+92,Plotnikov&Sironi19,Iwamoto+19}).  The ``synchrotron maser blastwave model" naturally accounts for the high observed linear polarization of many bursts (e.g.~\citealt{Ravi+16,Petroff+17}), their complex spectral energy distribution (e.g.~\citealt{Babul&Sironi20}), and the observed downward drifting of burst subpulses (e.g.~\citealt{Hessels+19,CHIME+19}) due to the decreasing Doppler factor as the blast wave decelerates as it sweeps up material \citep{Metzger+19,Margalit+20}.

Even among magnetized shock models, there exist variations depending on the nature of the upstream medium into which the ultra-relativistic flare ejecta from the magnetar collides.  \citet{Lyubarsky14} and \citet{Beloborodov17} postulate an upstream medium composed of electron/positron plasma, either representing a nebula surrounding the magnetar, or an ultra-relativistic spin-down powered wind, respectively.  \citet{Metzger+19} instead consider the upstream medium to be a mildly-relativistically expanding, baryon-loaded outflow with an electron-ion composition.  Such an ambient environment may be generated by prior flaring activity.  The existence of baryonic ejection is supported in at least the most powerful giant Galactic magnetar flares by radio afterglow emission indicating the presence of a trans-relativistic, mass-loaded outflow (\citealt{Frail+99,Gelfand+05}).  

As shown by \citet{Margalit+20b} and recounted below, the predictions of the baryon-loaded electron-ion upstream model agree quantitatively with the properties of \source, provided one interprets the hard X-ray emission from the flare \citep{Mereghetti+20,Zhang+20,Ridnaia+20,Tavani+20} as incoherent synchrotron radiation from relativistically-hot thermal electrons heated by the same shock generating the FRB.  Since both the radio and X-ray emission originate from the same physical location, their nearly-simultaneous observed arrival time$-$once accounting for the finite propagation speed of the radio burst through the Galactic ISM$-$is naturally expected  (see also \citealt{Yuan+20,Yamasaki+20}).  \citet{Margalit+20b} summarize several falsifiable predictions of this model, testable by future Galactic or nearby extragalactic FRBs (see also \citealt{Lu+20,Beniamini&Kumar20} for criticisms of shock models). 

An important consequence of an electron-ion upstream medium is the prediction at the shocks of heating, and potential non-thermal acceleration, of ions to relativistic energies.  Unlike an electron/positron plasma, relativistic ions can generate neutrino emission via the photohadronic interaction with thermal synchrotron photons, similar to proposed mechanisms of neutrino emission in gamma-ray burst jets \citep{Waxman&Bahcall97,Meszaros&Waxman01,Dermer&Atoyan03,Guetta&Granot03}.  

In this Letter, we estimate the properties of the neutrino bursts predicted to accompany FRBs in the synchrotron maser electron-ion blastwave scenario, and assess their detectability with the IceCube Neutrino Observatory.  Table \ref{tab:def} summarizes several definitions which may be useful to refer to throughout the text.

\section{Magnetized Relativistic Shocks}

Following \citet{Metzger+19}, we consider an ultra-relativistic blast wave generated by flare ejecta from the magnetar (e.g. a magnetically-dominated plasmoid; e.g.~\citealt{Yuan+20}) with an isotropic energy $\mathcal{E}_{\rm flare}$ propagating into an external electron-ion medium with a radial density profile $n_{\rm ext} \propto r^{-k}$, where $k < 3$ (we will soon specialize to the case $k = 0$).  We focus on observer times $t$ greater than the intrinsic width of the flare ejecta $\delta t$, as is typically satisfied for FRBs of duration $\gtrsim 1$ ms if the flare duration $\delta t \sim few \times R_{\rm ns}/c \sim 0.1$ ms is set by the Alfv\'en crossing time of the inner magnetosphere of a neutron star of radius $R_{\rm ns} \sim 10$ km.   

The deceleration of the blastwave follows a self-similar evolution \citep{Blandford&McKee76} resulting in the following evolution for the shock, Lorentz factor, and kinetic luminosity of the forward shock,
\begin{eqnarray}
\Gamma &=& \Gamma_{\rm FRB}\left(\frac{t}{t_{\rm FRB}}\right)^{\frac{(k-3)}{2(4-k)}} \underset{k=0}= \Gamma_{\rm FRB}\left(\frac{t}{t_{\rm FRB}}\right)^{-\frac{3}{8}}  ,\\
r_{\rm sh} &=& r_{\rm FRB}\left(\frac{t}{t_{\rm FRB}}\right)^{\frac{1}{(4-k)}} \underset{k=0}=r_{\rm FRB}\left(\frac{t}{t_{\rm FRB}}\right)^{\frac{1}{4}} , \\
n_{\rm ext} &=& n_{\rm FRB}\left(\frac{t}{t_{\rm FRB}}\right)^{\frac{-k}{(4-k)}} \underset{k=0}=n_{\rm FRB},\\ \nonumber \\
L_{\rm sh} &=& L_{\rm FRB}\left(\frac{t}{t_{\rm FRB}}\right)^{-1} \simeq  \frac{\mathcal{E}_{\rm flare}}{4 t_{\rm FRB}}\left(\frac{t}{t_{\rm FRB}}\right)^{-1} ,
\label{eq:evolution}
\end{eqnarray}
where $t_{\rm FRB}$ is the observed duration of the FRB (time for the spectral energy distribution of the maser emission to sweep across the observing band; \citealt{Metzger+19}) and $\{\Gamma_{\rm FRB}, r_{\rm FRB}, n_{\rm FRB},L_{\rm FRB}\}$ are the shock properties at $t \sim t_{\rm FRB}$.  

The latter quantities are related to observable properties of the FRB, namely its duration $t_{\rm FRB} = t_{\rm ms}$ ms, radio frequency $\nu_{\rm obs} = \nu_{\rm GHz}$ GHz, isotropic radio energy $\mathcal{E}_{\rm radio} = \mathcal{E}_{\rm r,40} 10^{40}$ erg, according to \citep{Margalit+20}
\begin{eqnarray}
r_{\rm FRB} \approx 5\times 10^{12}{\rm cm}\,\,f_{\xi,-3}^{-2/15}\nu_{\rm GHz}^{-7/15}t_{\rm ms}^{1/5}\mathcal{E}_{\rm r,40}^{1/3}, \\
n_{\rm FRB} \approx 420\,{\rm cm^{-3}}\,f_{\xi,-3}^{-4/15}\nu_{\rm GHz}^{31/15}t_{\rm ms}^{2/5}\mathcal{E}_{\rm r,40}^{-1/3} , \\
\Gamma_{\rm FRB} \approx 287 \,f_{\xi,-3}^{-1/15}\nu_{\rm GHz}^{-7/30}t_{\rm ms}^{-2/5}\mathcal{E}_{\rm r,40}^{1/6} \label{eq:GammaFRB}
\end{eqnarray}
where $f_{\xi} = 10^{-3}f_{\xi,-3}$ is the efficiency of the maser emission, normalized to a characteristic value found by particle-in-cell (PIC) plasma simulations of magnetized shocks \citep{Plotnikov&Sironi19,Babul&Sironi20} and we have assumed a upstream ratio of electrons to nucleons $f_e = 0.5$ (the results can be easily generalized to include an electron-positron component). 

The radiative efficiency of the FRB emission is given by
\begin{equation}
\eta \equiv \frac{\mathcal{E}_{\rm radio}}{\mathcal{E}_{\rm flare}} \approx 4\times 10^{-5}f_{\xi,-3}^{4/5}\nu_{\rm GHz}^{-1/5}t_{\rm ms}^{-1/5},
\label{eq:eta}
\end{equation}
such that 
\begin{equation}
\mathcal{E}_{\rm flare} \approx 2.4\times 10^{44}{\rm erg}\,f_{\xi,-3}^{-4/5}\nu_{\rm GHz}^{1/5}t_{\rm ms}^{1/5}\mathcal{E}_{\rm r,40}
\label{eq:Eflare}
\end{equation}
and the kinetic power of the shock at the epoch of FRB emission is given by
\begin{equation}
L_{\rm FRB} \simeq \frac{\mathcal{E}_{\rm flare}}{4 t_{\rm FRB}} \approx 6\times 10^{46}\,{\rm erg\,s^{-1}}f_{\xi,-3}^{-4/5}\nu_{\rm GHz}^{1/5}t_{\rm ms}^{-4/5}\mathcal{E}_{\rm r,40}
\end{equation}
The population of cosmological FRBs with known distances exhibit typical ranges $\mathcal{E}_{\rm radio} \sim 10^{37}-10^{41}$ erg, $t_{\rm FRB} \sim 0.1-10$ ms, such that one infers $r_{\rm FRB} \sim 10^{12}-10^{13}$ cm, $\Gamma_{\rm FRB} \sim 10^{2}-10^{3}$, and $n_{\rm FRB} \sim 10^{2}-10^{4}$ cm$^{-3}$ \citep{Margalit+20}.  From the observed rate of downward frequency drifting of the sub-pulses, which in this model could arise from deceleration of the blast wave \citep{Metzger+19}, one infers power-law indices of the external medium spanning a wide-range $k\approx [-2,2]$ \citep{Margalit+20}.  In what follows we assume a constant density medium ($k = 0$) for simplicity, though this too can be easily generalized and our results to follow do not depend qualitatively on this assumption.

From equations (\ref{eq:evolution}, \ref{eq:GammaFRB}), the blastwave transitions from ulta-relativistic to non-relativistic expansion ($\Gamma \simeq 1$) after a time
\begin{equation}
t_{\rm nr} \simeq t_{\rm FRB}\Gamma_{\rm FRB}^{8/3} \simeq 3.5\times 10^{3}{\rm s}\, \,f_{\xi,-3}^{-8/45}\nu_{\rm GHz}^{-28/45}t_{\rm ms}^{-1/15}\mathcal{E}_{\rm r,40}^{4/9},
\label{eq:tNR}
\end{equation}
which is minutes to hours for typical parameters.

In addition to generating the coherent FRB emission, the shock heats the plasma to ultra-relativistic temperatures.  Again guided by the results of PIC simulations of magnetized shocks,  we assume that the post-shock energy is shared roughly equally with the electrons and ions, heating each to a Maxwellian distribution of energies (e.g.~\citealt{Sironi&Spitkovsky11}).  Given the strong magnetic field behind the shock, the electrons radiate (incoherent) synchrotron emission.  The latter peaks at a photon energy \citep{Metzger+19}
\begin{equation}
\epsilon_{\rm pk} = \frac{\hbar eB'}{m_e c}\bar{\gamma}^{2}\Gamma,
\end{equation}
where $\bar{\gamma} = (m_p/m_e)\Gamma/2$ is the mean thermal Lorentz factor, $B' = \sqrt{64\pi \sigma \Gamma^{2}m_p c^{2}n_{\rm ext}}$ is the post-shock magnetic field, and $\sigma$ is the magnetization of the upstream medium.  Thus we have,
\begin{eqnarray}
\epsilon_{\rm pk} \approx 
235\,{\rm MeV} \sigma_{-1}^{1/2}f_{\xi,-3}^{-2/5}\nu_{\rm GHz}^{1/10}\mathcal{E}_{\rm r,40}^{1/2}t_{\rm ms}^{-7/5}\left(\frac{t}{t_{\rm FRB}}\right)^{-3/2},
\label{eq:Epk}
\end{eqnarray}
in the hard X-ray/$\gamma$-ray range, where $\sigma_{-1} \equiv \sigma/(0.1)$.\footnote{The value of $\sigma$ in the baryon-loaded upstream is uncertain.  A minimum value $\sigma \gtrsim 10^{-3}$ is required for operation of the synchrotron maser, while very large values $\sigma \gg 1$ are also disfavored due to the declining radiative efficiency with increasing $\sigma$ \citep{Plotnikov&Sironi19}.}  The cooling frequency is given by
\begin{eqnarray}
\epsilon_{\rm c} &=& \frac{\hbar eB'}{m_e c}\gamma_{\rm c}^{2}\Gamma \approx \nonumber \\
&&0.023\,{\rm MeV}\,\,\sigma_{-1}^{-3/2}f_{\xi,-3}^{2/3}\nu_{\rm GHz}^{-65/30}t_{\rm ms}^{-1}\mathcal{E}_{\rm r,40}^{-1/6}\left(\frac{t}{t_{\rm FRB}}\right)^{-1/2} ,
\label{eq:Ecool}
\end{eqnarray}
where $\gamma_{\rm c} = (6\pi m_e c/\sigma_{\rm T}\Gamma B'^{2}t)$ is the Lorentz factor of the electrons that cool on the expansion time.  The post-shock electrons thus remain fast-cooling ($\epsilon_{\rm c} \lesssim \epsilon_{\rm pk}$) for a timescale
\begin{eqnarray}
t_{\rm c} \approx 10\,{\rm s}\,\sigma_{-1}^{2}f_{\xi,-3}^{-16/15}\nu_{\rm GHz}^{34/15}t_{\rm ms}^{3/5}\mathcal{E}_{\rm r,40}^{2/3},
\label{eq:tc}
\end{eqnarray}
much longer than the FRB duration; thus the radiated X-ray/gamma-ray energy $E_{\rm X/\gamma} \approx \mathcal{E}_{\rm flare}/2$ is equal to that given to the electrons.

Considering the application to \source~\citep{Margalit+20}, the value $\eta \sim 10^{-4}-10^{-5}$ predicted by equation \ref{eq:eta} agrees with the measured ratio of the radio and X-ray fluence ($\mathcal{E}_{\rm radio} \sim 10^{34}-10^{35}$ erg; \citealt{CHIME+20,Bochenek+20}; $E_{\rm X/\gamma} \sim 10^{39}-10^{40}$ erg; \citealt{Mereghetti+20,Zhang+20}) for $f_{\xi} \sim 10^{-3}$.  For the same parameters, equation (\ref{eq:Epk}) predicts a peak frequency (given the observed duration $t \sim 3$ ms of the X-ray spikes of \source) of $\epsilon_{\rm pk} \sim 300\sigma_{-1}^{1/2}$ keV, also consistent with the spectral peak of the observed power-law X-ray/gamma-ray emission.  Finally, the energy spectrum of the X-ray/gamma-ray emission is well-fit by a power-law  + cut-off (e.g.~\citealt{Mereghetti+20}), similar to predictions for a fast-cooling thermal synchrotron spectrum \citep{Giannios&Spitkovsky09}.  
Taken together, these observations support a magnetized shock propagating into a baryon-loaded medium as the origin of \source~and its associated X-ray burst \citep{Margalit+20b}.  

\section{High-Energy Neutrino Emission}

\subsection{Proton Heating and Acceleration}

A fraction $\sim 1/2$ of the shock's energy goes into thermally heating ions, which we hereafter take to be protons for simplicity.  The protons achieve a relativistic Maxwellian distribution with an average particle energy of $\bar{E}' \approx \Gamma m_p c^{2}/2$ in the co-moving frame of the shocked gas (hereafter denoted by a prime ').   Though less certain, particularly for a strongly magnetized upstream ($\sigma \gg 1$), the shocks may accelerate a small fraction of the protons to higher energies $\gg \bar{E}' $ (e.g.~via diffusive acceleration; \citealt{BlandfordOstriker78}), into a power-law non-thermal spectrum of the form \begin{equation}
\frac{dN'_p}{dE'} \propto  \left(\frac{E'}{\bar{E}'}\right)^{-q}, \,\,\,  \bar{E}' < E'  < E_{max}',
\label{eq:Np}
\end{equation}
where $E_{max}'$ is a maximum cut-off energy, to be estimated below.  Taking $q \simeq 2$, the fraction of the total shock kinetic luminosity placed into protons of energy $\sim E' \ge \bar{E}'$ can be roughly written as,
\begin{equation} f_{E'} \approx \begin{cases}
1/2 & E \sim \bar{E}  \\
\epsilon_{\rm rel}/\Lambda & E \gg \bar{E} ,
\end{cases}
\label{eq:fEp}
\end{equation}
where $\epsilon_{\rm rel} \lesssim 0.1$ is an acceleration efficiency and $\Lambda \equiv {\rm ln}(E_{max}'/\bar{E}') \sim 10$ (see below).

In order to estimate the value of $E_{max}'$ we equate the proton acceleration time (estimated by the Larmor gyration time) $t_{\rm acc} \sim 2\pi E'/eB'c$ to various loss timescales.   The most stringent constraint\footnote{Other proton loss timescales, such as due to synchrotron cooling, are much longer$-$and hence unconstraining on $E_{max}-$ than the dynamical timescale for shock parameters of interest.} comes from equating $t_{\rm acc}$ to the dynamical time, $t_{\rm dyn} \simeq r_{\rm sh}/(\Gamma c)$, over which adiabatic losses occur.  This gives
\begin{eqnarray}
E_{max}' &\simeq& \frac{eB'r_{\rm sh}}{2\pi \Gamma} = er_{\rm sh}c\sqrt{(16/\pi)\sigma m_p n_{\rm ext}} \nonumber \\
&\approx& 850\,\,{\rm TeV}\sigma_{-1}^{1/2}f_{\xi,-3}^{-4/15}\nu_{\rm GHz}^{17/30}t_{\rm ms}^{2/5}\mathcal{E}_{\rm r,40}^{1/6}\left(\frac{t}{t_{\rm FRB}}\right)^{1/4}.
\label{eq:Emax}
\end{eqnarray}
Note that $E_{max}' \sim 1\, {\rm PeV} \sim 10^{3}-10^{4}\bar{E}'$ on the timescale of the FRB $t \sim t_{\rm FRB}$, increasing by $t \sim t_{\rm nr}$ to $\gtrsim 10$ PeV.


\subsection{Pion Production and Neutrino Emission}

Although the electrons generally cool faster than the dynamical timescale at early times ($t \ll t_{\rm c}$; eq.~\ref{eq:tc}), this is not true of the protons due to their greater mass.  The protons retain their energy, enabling them to interact with the thermal synchrotron photons via the photomeson process $p +\gamma \rightarrow p + \pi^{\pm,0}$ \citep{Berezinskii+90}, generating high-energy gamma-rays and neutrinos via the decay of $\pi^{0}$ and $\pi^{\pm}$, respectively.  The rate of hadronuclear interactions (e.g.~$p + p \rightarrow  p + p + \pi^{\pm,0}$) are low in comparison because of the high photon-to-baryon ratio behind the shock.

The energy distribution of the thermal synchrotron emission from the electrons peaks at a photon energy
\begin{eqnarray}
\epsilon_{\rm pk}' \approx \frac{\epsilon_{\rm pk}}{\Gamma} \approx
 0.8\,{\rm MeV}\sigma_{-1}^{1/2}f_{\xi,-3}^{-1/3}\nu_{\rm GHz}^{1/3}t_{\rm ms}^{-1}\mathcal{E}_{\rm r,40}^{1/3}\left(\frac{t}{t_{\rm FRB}}\right)^{-9/8} ,
\end{eqnarray}
where we have used equations (\ref{eq:GammaFRB}) and (\ref{eq:Epk}).  Immersed in this radiation field, pion production peaks at the $\Delta$-resonance at a proton energy (e.g.~\citealt{Waxman&Bahcall97})
\begin{eqnarray}
&&E_{\rm thr}' \approx \frac{E_{\Delta}'}{\epsilon_{\rm pk}'}\frac{m_p c^{2}}{2} \approx \nonumber \\
 &&170\,{\rm GeV} \sigma_{-1}^{-1/2}f_{\xi,-3}^{1/3}\nu_{\rm GHz}^{-1/3}t_{\rm ms}\mathcal{E}_{\rm r,40}^{-1/3}\left(\frac{t}{t_{\rm FRB}}\right)^{9/8},
\end{eqnarray}
where $E_{\Delta}' \approx 0.3$ GeV.  Comparing this threshold to the proton thermal energy,
\begin{eqnarray}
&&\frac{E_{\rm thr}'}{\bar{E}'} \approx \frac{2 E_{\rm thr}'}{\Gamma m_p c^{2}}  \approx \frac{ E_{\Delta}'}{\epsilon_{\rm pk}} \nonumber \\
 && \approx 1.3 \,\sigma_{-1}^{-1/2}f_{\xi,-3}^{2/5}\nu_{\rm GHz}^{-1/10}t_{\rm ms}^{7/5}\mathcal{E}_{\rm r,40}^{-1/2}\left(\frac{t}{t_{\rm FRB}}\right)^{3/2}.
\label{eq:where}
\end{eqnarray}
Thus, on timescales of the FRB emission ($t \sim t_{\rm FRB})$, protons which are either thermal ($E' \sim \bar{E}'$), or only moderately into the non-thermal tail ($E' \gtrsim \bar{E}'$; eq.~\ref{eq:Np}), are capable of pion-producing on synchrotron photons.  However, at times $t \gg t_{\rm FRB}$, only the non-thermal protons with $E' \gg \bar{E}'$ have enough energy to produce pions.
This transition ($\bar{E}' = E_{\rm thr}'$) from thermal to non-thermal-only production occurs at a time
\begin{equation}
t_{\rm nth} \approx 8.5\times 10^{-4}{\rm s}\,\,\sigma_{-1}^{1/3}f_{\xi,-3}^{-4/15}\nu_{\rm GHz}^{1/15}t_{\rm ms}^{1/15}\mathcal{E}_{\rm r,40}^{1/3} .
\label{eq:tnth}
\end{equation}

The pions generated decay into neutrinos of lab-frame energy $\varepsilon \approx \Gamma E'/20$  (\citealt{Kelner&Aharonian08}).  The minimum neutrino energy at time $t$ is therefore   
\begin{eqnarray}
&& \varepsilon_{\rm min} \approx {\rm max}[\Gamma \bar{E}'/20,\Gamma E_{\rm thr}'/20] \approx \nonumber \\
&& {\rm max} \begin{cases} 
1.9\,{\rm TeV} f_{\xi,-3}^{-2/15}\nu_{\rm GHz}^{-7/15}t_{\rm ms}^{-4/5}\mathcal{E}_{\rm r,40}^{1/3}\left(\frac{t}{t_{\rm FRB}}\right)^{-3/4}, t \lesssim t_{\rm nth}\\
2.5\,{\rm TeV} \sigma_{-1}^{-1/2}f_{\xi,-3}^{4/15}\nu_{\rm GHz}^{-17/30}t_{\rm ms}^{3/5}\mathcal{E}_{\rm r,40}^{-1/6}\left(\frac{t}{t_{\rm FRB}}\right)^{3/4}, t \gtrsim t_{\rm nth} \nonumber \\
\end{cases} \\
\label{eq:Enumin}
\end{eqnarray}%

From equation (\ref{eq:Emax}), one can also define an absolute maximum neutrino energy, as set by the accelerator,
\begin{eqnarray}
\varepsilon_{\rm max} \approx \frac{\Gamma E_{\rm max}'}{20} 
 \approx 12\,{\rm PeV}\,\,\sigma_{-1}^{1/2}f_{\xi,-3}^{-1/3}\nu_{\rm GHz}^{1/3}\mathcal{E}_{\rm r,40}^{1/3}\left(\frac{t}{t_{\rm FRB}}\right)^{-1/8}
\end{eqnarray}
Equating $\varepsilon_{\rm min}(t > t_{\rm nth})$ (eq.~\ref{eq:Enumin}) to $\varepsilon_{\rm max}$, we obtain
\begin{equation}
t_{\rm max} \approx 17\,{\rm s}\,\,\sigma_{-1}^{8/7}f_{\xi,-3}^{-24/35}\nu_{\rm GHz}^{36/35}t_{\rm ms}^{11/35}\mathcal{E}_{\rm r,40}^{4/7}.
\label{eq:tmax}
\end{equation}
This is the maximal timescale for significant neutrino emissions because at $t > t_{\rm max}$ the proton energy required to interact with the most energetic synchrotron photons (of energy $\sim \epsilon_{\rm pk}$) exceeds $E_{\rm max}$.

\begin{deluxetable}{cc}
\tablecaption{Definitions\label{tab:def}}
\tablewidth{700pt}
\tabletypesize{\scriptsize}
\tablehead{
\colhead{Symbol} & \colhead{Description} 
} 
\startdata
$E$ & Proton energy \\
$\epsilon$ & Photon energy \\
$\varepsilon$ & Neutrino energy \\
$\mathcal{E}$ & Total energy \\
\enddata
\end{deluxetable}

\begin{deluxetable}{ccc}
\tablecaption{Key Timescales\label{tab:timescales}}
\tablewidth{700pt}
\tabletypesize{\scriptsize}
\tablehead{
\colhead{Symbol} & \colhead{Description} & 
\colhead{Typical Value$^{\dagger}$} 
} 
\startdata
$t_{\rm FRB}$ & FRB duration & $\sim 0.1-10$ ms \\
$t_{\rm nth}$ (eq.~\ref{eq:tnth}) & Duration of $\nu$'s from thermal protons & $\sim 0.1-1$ ms \\
$t_{\rm c}$ (eq.~\ref{eq:tc}) & Synchrotron emission no longer fast-cooling  & $\sim 0.1-10^{3}$ s \\ $t_{\rm max}$ (eq.~\ref{eq:tmax}) & Duration of $\nu$'s from non-thermal protons & $\sim 1-10^{3}$ s \\ 
$t_{\rm nr}$ (eq.~\ref{eq:tNR}) & Transition to non-relativistic shock & $\sim 10^{2}-10^{5}$ s \\
\enddata
$^{\dagger}$For $\sigma \sim 0.1$ and $\mathcal{E}_{\rm rad} \sim 10^{37}-10^{43}$ erg in the range of cosmological FRB.\\
\end{deluxetable}

Before $t_{\rm max}$, higher-energy neutrinos with $\varepsilon_{\rm min} < \varepsilon < \varepsilon_{\rm max}$ can be created by the interaction of protons with energies $E' \gg E_{\rm thr}', \bar{E}'$ with photons of lower energies $\epsilon' \ll \epsilon_{\rm pk}'$ below the synchrotron peak.  In fact, at times $t < t_{\rm c}$ (eq.~\ref{eq:tc}) such interactions dominate neutrino production (by non-thermal protons) because the photon number spectrum in the fast-cooling regime,
\begin{equation}
\epsilon(dN_{\gamma}/d\epsilon) \propto \epsilon^{-1/2}, \epsilon_{\rm c} \lesssim \epsilon \lesssim \epsilon_{\rm pk},
\label{eq:spectrum}
\end{equation}
peaks at $\epsilon_{\rm c} < \epsilon_{\rm pk}$.\footnote{At times $t \gtrsim t_{\rm c}$ the spectrum below $\epsilon_{\rm pk}$ steepens to the standard slow-cooling spectrum $\epsilon(dN_{\gamma}/d\epsilon) \propto \epsilon^{1/3}$ for which photons of energy $\epsilon_{\rm pk}$ dominate by both energy {\it and} number.} 

Before $t_{\rm c}$ the energy distribution of neutrinos from non-thermal protons will thus peak at an energy,
\begin{eqnarray}
\varepsilon_{\rm c} &\approx& \frac{\Gamma E_{\Delta}'}{20 \epsilon_{\rm c}'}\frac{m_p c^{2}}{2}
 \approx \nonumber \\
 &26\,{\rm PeV}&\,\sigma_{-1}^{3/2}f_{\xi,-3}^{-4/5}\nu_{\rm GHz}^{17/10}t_{\rm ms}^{1/5}\mathcal{E}_{\rm r,40}^{1/2}\left(\frac{t}{t_{\rm FRB}}\right)^{-1/4},
\label{eq:Ec}
\end{eqnarray}
where $\epsilon_{\rm c}' = \epsilon_{\rm c}/\Gamma$.  The value of $\varepsilon_{\rm c}$ decreases in time, matching $\varepsilon_{\rm min}$ at $t = t_{\rm c}$.  

In principle, pions and muons could suffer synchrotron losses before decaying into neutrinos.  This happens above a energy critical energy 
\begin{eqnarray}
    \varepsilon_{\pi,\mu}^{\rm syn} &=& \left(\frac{6\pi m_{\pi,\mu} c}{\sigma_T B'^2 \tau_{\pi,\mu}}\right)^{1/2} \left(\frac{m_{\pi,\mu}}{m_e}\right) m_{\pi,\mu}\,c^2\Gamma \\ \nonumber
    &=& (30, \, 1.7)\times 10^{18}{\,\rm eV}\,\sigma_{-1}^{-1/2}f_{\xi,-3}^{2/15}\nu_{\rm GHz}^{-31/30}t_{\rm ms}^{-1/5}\mathcal{E}_{\rm r,40}^{1/6},
\end{eqnarray}
where $m_{\pi,\mu}$ and $\tau_{\pi,\mu}$ are the masses and rest-frame lifetimes of pions and muons, respectively. Because $\varepsilon_{\pi}^{\rm syn}\gg 4\,\varepsilon_{\rm max}$ and $\varepsilon_{\mu}^{\rm syn}\gg 3\,\varepsilon_{\rm max}$ (where the prefactors account for the rough energy partition between pions/muons/neutrinos in the decays), we conclude that synchrotron cooling of the secondaries is negligible.

The timescales discussed in this section are summarized in Table \ref{tab:timescales}.  Although all of the timescales increase with the flare energy, the hierarchy $t_{\rm nth} < t_{\rm c} < t_{\rm max} < t_{\rm nr}$ is generally preserved for $\sigma \lesssim 1$ and moderate variations around fiducial parameters.

In summary, at early times ($t \lesssim t_{\rm  nth}$) we expect $\sim 1-10$ TeV neutrinos generated by the interaction of the thermal protons with synchrotron photons of energy $\sim \epsilon_{\rm pk}$.  This prediction is relatively robust (within the baryon shock model) because proton heating at the shocks is inevitable and since $t_{\rm nth} \sim t_{\rm FRB}$ does not need to assume a continuation of the external medium outside the region probed by the observed FRB.

At later times ($t \gtrsim t_{\rm nth}$) the thermal protons are soon no longer sufficiently energetic to produce pions, but nevertheless neutrinos may still be produced by the power-law non-thermal protons.  Insofar as the proton spectrum is flat ($q \simeq 2$) the dominant interaction will initially ($t \ll t_{\rm c}$) be between photons of energy $\sim \epsilon_{\rm c} \ll \epsilon_{\rm pk}$ on high-energy protons $\sim 20\varepsilon_{\rm c}$ generating neutrinos of energy $\varepsilon_{\rm c} \sim 1-10$ PeV.  However, at times $t \gg t_{\rm c}$ the synchrotron spectrum becomes slow-cooling and the interaction will be dominated by photons of energy $\sim \epsilon_{\rm pk}$ generating neutrinos of energy $\varepsilon_{\rm min}(t \gtrsim t_{\rm c}) \gtrsim 10$ PeV.   

\subsection{Neutrino Radiative Efficiency}

\begin{figure}
    \centering
    \includegraphics[width=0.5\textwidth]{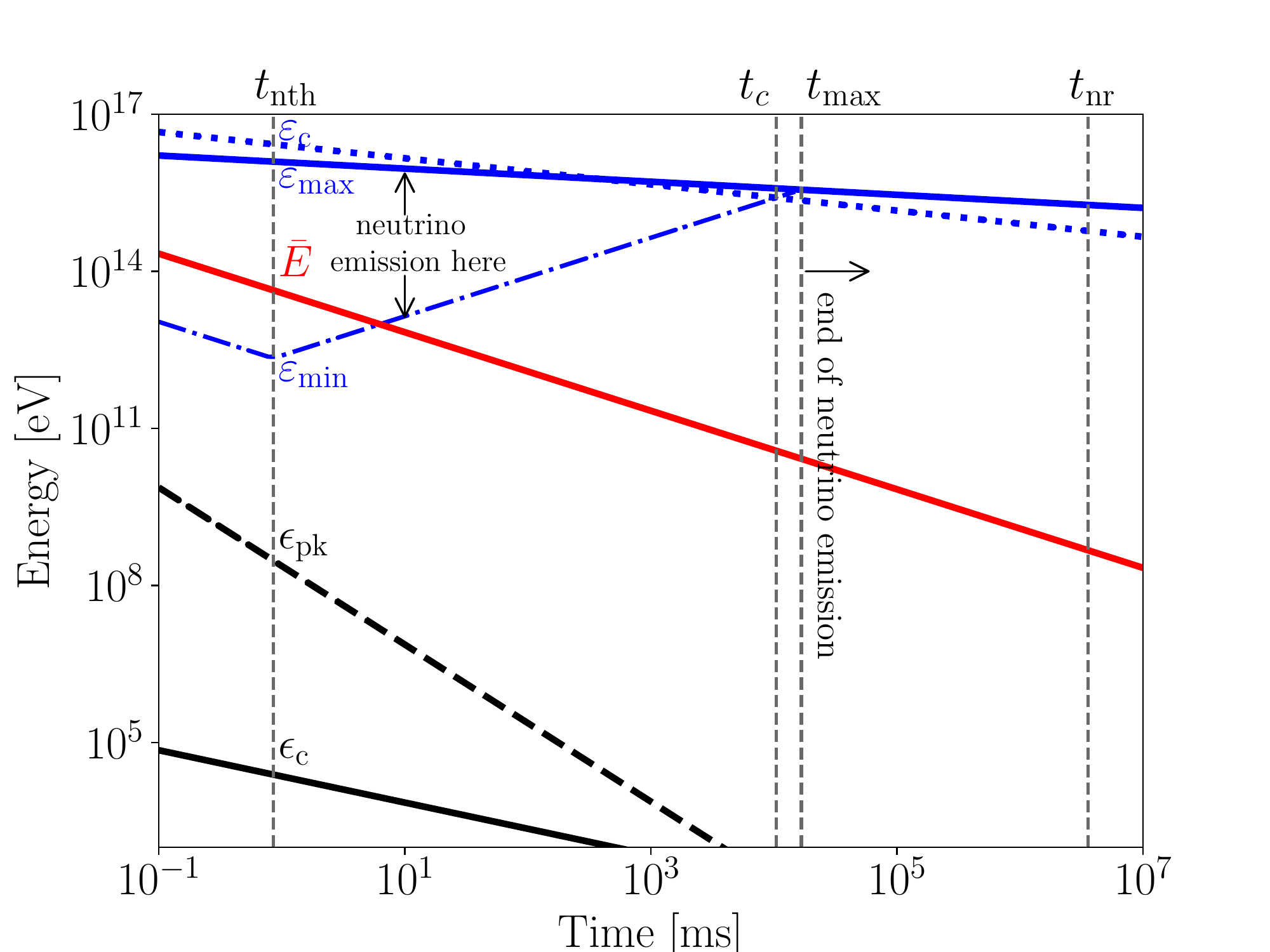}
    \includegraphics[width=0.5\textwidth]{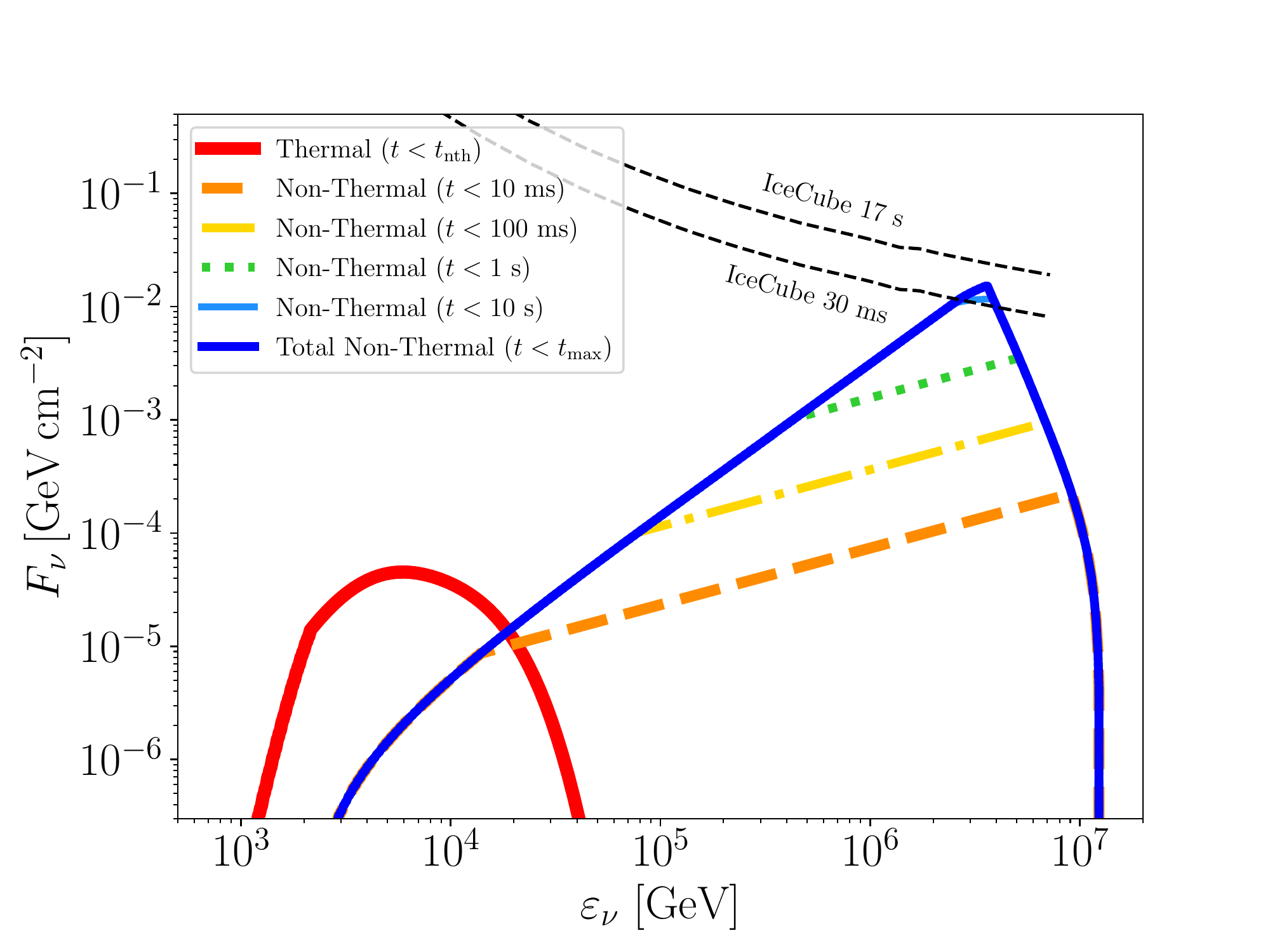}
    \caption{{\it Top panel:} Critical energies as a function of time for a fiducial model with $t_{\rm FRB} = 1$ ms, $\sigma = 0.1$, $f_{\xi} = 10^{-3}$, $\nu = 1$ GHz, $\mathcal{E}_{\rm radio} = 10^{40}$ erg.  {\it Bottom panel:} Observed muon neutrino spectra for the same model in the top panel, showing the growth of the spectra over decades in time as the shock propagates outwards.  Shown for comparison are approximate IceCube sensitivity curves for observation windows of 30 ms and 17 s, corresponding roughly to the duration of neutrino emission from thermal and non-thermal protons, respectively, for a source distance of 0.1 kpc.  }
    \label{fig:model}
\end{figure}

What fraction of the available proton luminosity $\sim L_{\rm sh}/2$ will be radiated as neutrinos?  One can define an ``optical depth" for neutrino production over each decade in observer time $\sim t$,
\begin{eqnarray}
\tau_{p\gamma} &\approx& n_{\gamma}'\sigma_{p \gamma}\kappa_{p\gamma} r'  
\end{eqnarray}
where $n_{\gamma}'$ is the target photon number density given below, $r' \approx r_{\rm sh}/\Gamma$ is the characteristic radius of the post-shock region in the comoving frame (through which relativistic protons accelerated at the shock traverse to escape without losses), $\sigma_{p\gamma}$ is the cross section of photopion production, which is $\sim 5\times 10^{-28}$ cm$^{2}$ \citep{Eidelman+04} corresponding to its value near the $\Delta$-resonance (and $\sim 1.6\times 10^{-28}$ cm$^{2}$ above the resonance).  The factor $\kappa_{p\gamma} \sim 0.15$ is the average fraction of energy lost from a proton per collision (``elasticity").  
Consider first proton interactions with photons of energy $\epsilon = \epsilon_{\rm pk}$ which generate neutrinos of energy $\varepsilon_{\rm min}$ (eq.~\ref{eq:Enumin}).  Their number density is given by $n_{\gamma}' \approx U_\gamma'/\epsilon_{\rm pk}'$, where
\begin{equation} U_{\gamma}' \approx \frac{L_{\rm sh}/2}{4\pi r_{\rm sh}^{2} c \Gamma^{2}}  \begin{cases}
1 & t \le t_{\rm c}  \\
\left(t/t_{\rm c}\right)^{-1/2}  & t \ge t_{\rm c} ,
\end{cases}
\end{equation}
where the factor of $1/2$ in the numerator arises from the half of the shock-power placed into electrons in the fast-cooling regime.  The factor $(t/t_{\rm c})^{-1/2}$ factor accounts for the reduction in the radiated power when the electrons become slow-cooling at $t \gtrsim t_{\rm c}$ (eq.~\ref{eq:tc}).

Combining with the above
\begin{eqnarray}
\tau_{p\gamma}[\varepsilon_{\rm \nu,min}] &\approx& \frac{L_{\rm sh}\sigma_{p\gamma}\kappa_{p\gamma}}{8\pi r_{\rm sh} c \Gamma^{2}\epsilon_{\rm pk}} \begin{cases}
1 & t \le t_{\rm c}  \\
(t/t_{\rm c})^{-1/2}  & t \ge t_{\rm c} ,
\end{cases} \nonumber \\
&\approx& 3.9\times 10^{-8}\sigma_{-1}^{-1/2}f_{\xi,-3}^{-2/15}\nu_{\rm GHz}^{31/30}t_{\rm ms}^{6/5}\mathcal{E}_{\rm r,40}^{-1/6}\times \nonumber \\
&&
\begin{cases}
\left(\frac{t}{t_{\rm FRB}}\right) & t \le t_{\rm c}  \\
\left(\frac{t_{\rm c}}{t_{\rm FRB}}\right)\left(\frac{t}{t_{\rm c}}\right)^{1/2}  & t \ge t_{\rm c}
\end{cases}
\end{eqnarray}
At time $t \lesssim t_{\rm c}$, the photon density $n_{\gamma}$ at $\epsilon = \epsilon_{\rm c}$ is larger by a factor $(t/t_{\rm c})^{-1/2}$ than at $\epsilon_{\rm pk}$ (eq.~\ref{eq:spectrum}); thus, the optical depth to generating $\varepsilon_{\rm \nu,c}$ neutrinos is larger by the same factor:
\begin{eqnarray}
&& \tau_{p\gamma}[\varepsilon_{\rm \nu,c}] \approx \tau_{p\gamma}[\varepsilon_{\rm \nu,min}](t/t_{\rm c})^{-1/2} \nonumber \\
&&\approx 3.9\times 10^{-6}\sigma_{-1}^{1/2}f_{\xi,-3}^{-2/3}\nu_{\rm GHz}^{65/30}t_{\rm ms}\mathcal{E}_{\rm r,40}^{1/6} \left(\frac{t}{t_{\rm FRB}}\right)^{1/2}, t \le t_{\rm c} \nonumber \\
\end{eqnarray}
Interestingly, $\tau_{\rm p,\gamma}$ at the peak of the neutrino energy spectrum increases $\propto t^{1/2}$ both before and after $t_{\rm c}$.  The fluence emitted in neutrinos over a timescale $\sim t$ is thus given by
\begin{eqnarray}
\mathcal{E}_{\nu} &\approx& \frac{3}{8}\int^{t} \tau_{p\gamma}f_{\rm E'}L_{\rm sh} dt \sim \frac{3}{8}\tau_{p\gamma}f_{\rm E'}L_{\rm sh}t \nonumber \\
&\approx& 8.9\times 10^{37}{\rm erg}\, f_{\rm E'} \sigma_{-1}^{1/2}f_{\xi,-3}^{-22/15}\nu_{\rm GHz}^{71/30}t_{\rm ms}^{6/5}\mathcal{E}_{\rm r,40}^{7/6}\left(\frac{t}{t_{\rm FRB}}\right)^{1/2}
\end{eqnarray}
where the factor $3/8$ accounts for the fraction of the proton energy that goes into neutrino products \citep{1999PhRvD..59b3002W} and $f_{\rm E'} \sim 0.01-0.5$ (eq.~\ref{eq:fEp}) is the fraction of the shock power placed into protons obeying the threshold condition.

One can define a neutrino radiative efficiency, 
\begin{eqnarray}
\eta_{\nu} &\equiv& \frac{\mathcal{E}_{\nu}}{\mathcal{E}_{\rm flare}} \approx \frac{3}{32} \tau_{p\gamma}[\varepsilon_{\rm \nu,c}] f_{\rm E'} \nonumber \\
 &\approx& 
3.7\times 10^{-7}f_{\rm E'}\sigma_{-1}^{1/2}f_{\xi,-3}^{-2/3}\nu_{\rm GHz}^{65/30}t_{\rm ms}\mathcal{E}_{\rm r,40}^{1/6}\left(\frac{t}{t_{\rm FRB}}\right)^{1/2} 
\end{eqnarray} 
which is valid at all times $t_{\rm FRB} \lesssim t \lesssim t_{\rm max}$. 

During the early epoch $t \lesssim t_{\rm nth}$, when neutrinos are produced by thermal protons, we have $f_{\rm E'} \sim 0.5$ (eq.~\ref{eq:fEp}) and hence $\eta_{\nu} \sim 10^{-7}-10^{-6}$ for fiducial parameters.  At times $t \gg t_{\rm nth} \sim t_{\rm FRB}$ when only non-thermal protons can produce neutrinos, we have $f_{\rm E'} \sim 10^{-2}$ if the shocks place a fraction $\epsilon_{\rm rel} \sim 0.1$ of their energy into the power-law tail.  Although the efficiency starts low in this case, it grows in time $\eta_{\nu} \propto t^{1/2}$ until reaching a peak value once $t \sim t_{\rm max}$ of 
\begin{eqnarray}
\eta_{\nu}[t_{\rm max}] \approx 4.8\times 10^{-7}\left(\frac{f_{\rm E'}}{10^{-2}}\right)
\sigma_{-1}^{\frac{15}{14}}f_{\xi,-3}^{-\frac{106}{105}}\nu_{\rm GHz}^{\frac{563}{210}}t_{\rm ms}^{\frac{23}{35}}\mathcal{E}_{\rm r,40}^{\frac{19}{42}}, 
\label{eq:etanu}
\end{eqnarray}
corresponding to a total radiated neutrino energy 
\begin{eqnarray}
&& {\cal E}_{\nu} = \eta_{\nu}[t_{\rm max}] \mathcal{E}_{\rm flare} \approx \nonumber \\
 && 1.1\times 10^{38}{\rm erg}\left(\frac{f_{\rm E'}}{10^{-2}}\right)\sigma_{-1}^{15/14}f_{\xi,-3}^{-38/21}\nu_{\rm GHz}^{121/42}t_{\rm ms}^{6/7}\mathcal{E}_{\rm r,40}^{61/42} \nonumber \\
 \label{eq:Eradnu}
\end{eqnarray}
The total neutrino emission from the shocks peaks at an energy,
\begin{equation}
\varepsilon_{\rm max}[t_{\rm max}] \approx 
3.6\,{\rm PeV}\,\,\sigma_{-1}^{5/14}f_{\xi,-3}^{-26/105}\nu_{\rm GHz}^{43/210}\mathcal{E}_{\rm r,40}^{11/42} t_{\rm ms}^{3/35} 
\label{eq:Emaxtmax}
\end{equation}
which scales weakly with FRB properties and model parameters.

\begin{figure}
    \centering
    \includegraphics[width=0.5\textwidth]{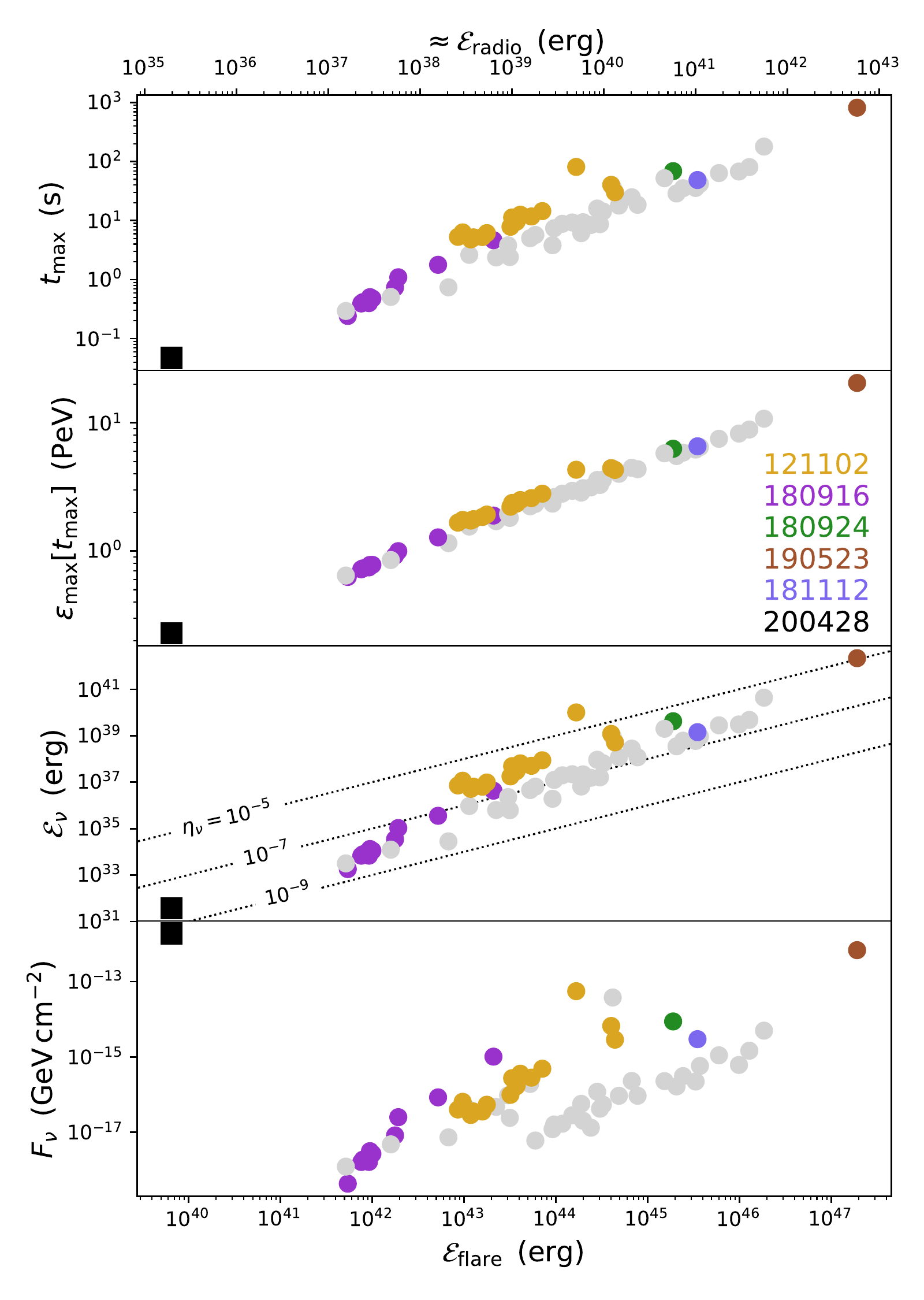}
    \caption{Predicted neutrino properties for a sample of FRBs with measured properties and known distances, as a function of the total flare energy ({\it bottom axis}) or approximate radio isotropic energy ({\it top axis}, assuming for simplicity a fixed radio efficiency $\eta = 3\times 10^{-5}$).  Properties shown include the total radiated energy in neutrinos $\mathcal{E}_{\nu}$ (eq.~\ref{eq:Eradnu}), duration of neutrino emission $t_{\rm max}$ (eq.~\ref{eq:tmax}), peak neutrino energy $\varepsilon_{\rm max}(t_{\rm max})$ (eq.~\ref{eq:Emaxtmax}), and neutrino flux $F_\nu$.  The FRBs shown include the sample of localized FRBs with known distances (different colors, as labeled) and non-localized CHIME FRBs (gray points), as in \cite{Margalit+20,Margalit+20b}.}
    \label{fig:FRB_population}
\end{figure}

Figure \ref{fig:model} shows an example calculation of the characteristic energies versus time (top panel) and the build-up of the neutrino spectrum with time as the shock propagates outwards (bottom panel).  Figure \ref{fig:FRB_population} shows several key properties related to neutrino emission$-$such as the duration, spectral peak, total radiated energy, and fluence$-$for a sample of FRBs with measured properties ($t_{\rm FRB}$, $\mathcal{E}_{\rm radio}$, $\nu_{\rm obs}$) and known distances.  Approximate IceCube sensitivity curves\footnote{These are oobtained by scaling the max-burst sensitivity of \citealt{2020ApJ...890..111A} for a power-law spectrum $E^{-2}$ with the detector effective area to northern-sky events.}  are shown for comparison.

\section{Detection Prospects}

The upshot of our work is the prediction of an efficiency of up to $\eta_{\nu} \sim 10^{-8}-10^{-5}$ (eq.~\ref{eq:etanu}; Fig.~\ref{fig:FRB_population}) for converting magnetar flare energy into a neutrino burst of energy $\sim 1-10$ PeV, near the peak sensitivity range of IceCube.  The energy radiated in neutrinos is typically $\sim 10-100$ times smaller than the radio isotropic energy, making the signal extremely challenging to detect.

Consider first the recent \source~from the Galactic magnetar SGR 1935+2154.  Taking $\mathcal{E}_{\rm flare} \sim 2E_{\rm X/\gamma} \sim 10^{40}$ erg, we predict a neutrino burst of energy $\mathcal{E}_{\nu} \lesssim 10^{32}$ erg, corresponding to a fluence (given an assumed source distance of 10 kpc) of $F_{\nu} \sim 10^{-12}$ GeV cm$^{-2}$ (Fig.~\ref{fig:FRB_population}).  This is $\sim 11$ orders of magnitude smaller than the IceCube upper limit of $5\times 10^{-2}$ GeV cm$^{-2}$ \citep{Vandenbroucke+20}.  Thus, the neutrino non-detection of FRB 200428 is not constraining on the baryon shock model.  

What about a more energetic Galactic magnetar flare?  Although no FRB was detected in association with the 2004 giant flare from SGR 1806-20 \citep{Tendulkar+16}, this could be explained if the shock-generated plasmoid ejection was directed away from our line of sight (in which case the FRB would be dimmer due to relativistic beaming effects).  Assuming a more fortuitious geometry of a future flare and scaling $\mathcal{E}_{\rm flare}$ up by a factor of $\sim 10^{6}$ to the $\sim 10^{46}$ erg energy scale of the gamma-ray emission from giant flares, the predicted flux for the nearest magnetar (1E 1048.1-5937 at $\approx 2.7$ kpc; \citealt{Gaensler+05}) could reach $\sim 10^{-2}$ GeV cm$^{-2}$.  In this case the detection of a neutrino from a giant flare becomes potentially feasible (see also \citealt{Gelfand+05}), particularly given proposed future upgrades to neutrino detectors (e.g.~\citealt{2020arXiv200804323T}).  Note that an FRB energy $\gtrsim 10^{12}$ Jy$\cdot$ ms is thus required to produce detectable neutrino emission in this case.

Individual extragalactic FRBs discovered to date are unlikely to be detectable due to their much greater distances (bottom panel of Fig.~\ref{fig:FRB_population}).  However, thanks to new surveys like ASKAP (\citealt{Bannister+19}) and CHIME/FRB (\citealt{CHIME+18}), the rate of FRB discoveries is rapidly growing and may reach thousands or more over the coming years.  

The distribution of FRB energies $dN/d\mathcal{E}_{\rm radio} \propto \mathcal{E}_{\rm radio}^{-\gamma}$ is roughly flat or increasing with energy ($\gamma \lesssim 2$; \citealt{Law+17,Lu&Piro19,Cruces+20}; though see e.g.  \citealt{Gourdji+19}), i.e. the total radiated energy is dominated by the rarest, most energetic bursts.  Given the neutrino radiative efficiency of the shocks $\eta_{\nu} \propto \mathcal{E}_{\rm radio}^{19/42}$ (eq.~\ref{eq:etanu}), the total radiated neutrino energy is also dominated by the high-$\mathcal{E}_{\nu}$ bursts.

A key question then becomes the upper cut-off energy to the FRB energy distribution, $\mathcal{E}_{\rm r,max}$.  Modeling the ASKAP FRB sample, \citet{Lu&Piro19} find a range of allowed values $\mathcal{E}_{\rm r,max} \approx 3\times 10^{42}-2\times 10^{44}$ erg for a burst frequency width $\nu \sim 1.4$ GHz.  In what follows, we optimistically assume\footnote{A burst of energy $\mathcal{E}_{\rm r,max} \sim 10^{44}$ erg corresponds to a flare of energy $\mathcal{E}_{\rm flare} \sim 3\times 10^{48}$ erg (eq.~\ref{eq:Eflare}), approaching the entire magnetic energy budget of a magnetar.} $\mathcal{E}_{\rm r,max} \approx 10^{44}$ erg, bursts for which the local volumetric rate is estimated as $\mathcal{R}(z = 0) \sim 6$ Gpc$^{-3}$ yr$^{-1}$ (adopting a $\gamma = 1.6$ power-law luminosity function; \citealt{Lu&Piro19,Luo+20}).  At this rate, the closest FRB of this energy per year occurs at a distance $D_{\rm min} \sim$ 0.34 Gpc.  Given its predicted neutrino energy $\mathcal{E}_{\nu} \sim 5\times 10^{43}$ erg (eq.~\ref{eq:Eradnu}), its neutrino flux would be $F_{\nu} \approx \mathcal{E}_{\nu}/(4\pi D_{\rm min}^{2}) \sim 10^{-9}$ GeV cm$^{-2}$, i.e. still $\sim$7 orders of magnitude below the IceCube detection threshold.  

The sensitivity could be improved by a stacked joint analysis of a large FRB sample (e.g.~\citealt{2018ApJ...857..117A, Kheirandish+19}).  Unfortunately, given the relatively long duration $t_{\rm max} \sim 3000$ s for the most energetic $\mathcal{E}_{\rm radio} \approx 10^{44}$ erg bursts described above,  the search becomes background-dominated after only a few bursts, after which the detector sensitivity quickly saturates. 
Although less energetic FRBs are more common and produce shorter-lived neutrino bursts $t_{\rm max}\propto \mathcal{E}_{\rm radio}^{4/7}$, their sharply lower radiated energies $\mathcal{E}_{\nu} \propto \mathcal{E}_{\rm radio}^{61/42}$ largely cancel out these benefits.    

\begin{figure}
    \centering
    \includegraphics[width=0.5\textwidth]{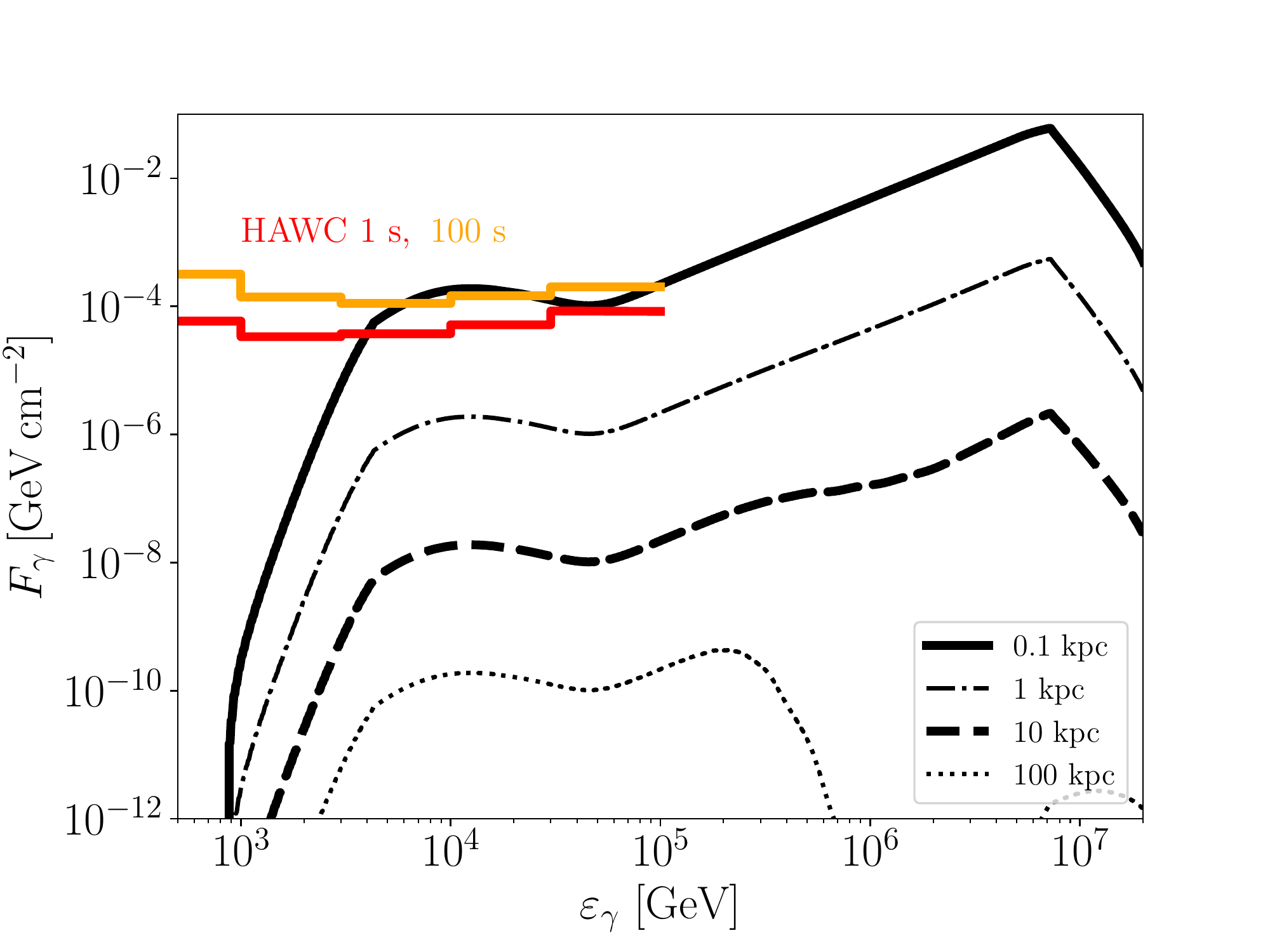}
    \caption{Total $\gamma$-ray fluence due to neutral pion decay from a fiducial burst as the one shown in Figure~\ref{fig:model} at a distance of 0.1, 1, 10 and 100~kpc, comparing to the sensitivities of the HAWC Observatory at $0^\circ$ zenith angle \citep{2019arXiv190806122M}.    $\gamma$-rays above 100~TeV are attenuated by pair production on the extragalactic background light after propagating over a distance of tens of kiloparsecs. }
    \label{fig:gammarays}
\end{figure}

In addition to a burst of neutrinos (from $\pi^{\pm}$ decay), we predict a burst of $\sim$ TeV$-$PeV gamma-rays from neutral pion decay.  As shown in Figure \ref{fig:gammarays}, the TeV thermal neutrinos fall in the sensitivity range of ground-based water Cerenkov detectors, including  HAWC \citep{2019arXiv190806122M}, and upcoming and future detectors LHAASO \citep{2016NPPP..279..166D} and SWGO \citep{2019BAAS...51g.109H}.  However, again for fiducial parameters we require a nearby source $\lesssim 0.1$ kpc for a detection.  Furthermore, the pion decay signal could be overwhelmed by non-thermal leptonic emission, depending on the electron acceleration efficiency of the shocks.  

\section{Conclusions}

With the discovery of \source, flaring magnetars are the leading FRB model.  However, several distinct mechanisms and environments around a magnetar for generating the radio burst$-$and any accompanying higher-frequency afterglow$-$have been proposed, which have proven challenging to distinguish observationally.  A unique prediction of baryon-loaded shock-powered FRB models \citep{Metzger+19,Margalit+20} are neutrinos generated by ions heated or accelerated at the shock interact with thermal synchrotron shock photons via the photo-hadronic process.  

We predict a burst of $\sim$ TeV$-$PeV neutrinos of total energy $\mathcal{E}_{\nu} \approx 10^{35}-10^{44}$ erg (depending most sensitively on the FRB isotropic energy; eq.~\ref{eq:Eradnu}) which lasts for a timescale $t_{\rm max} \sim 0.1-1000$ s (eq.~\ref{eq:tmax}) following the radio burst (once accounting for the time delay due to the finite propagation speed of the radio waves as inferred from the dispersion measure).  Our calculations make several optimistic assumptions, including (1) non-thermal particle acceleration at magnetized shocks with an efficiency of $10\%$; (2) the presence of a baryon-loaded medium which extends at least a distance $r_{\rm max} \equiv r[t_{\rm max}] \sim (3-30)r_{\rm FRB}$ beyond the radius $r_{\rm FRB}$ probed by the FRB emission itself.\footnote{We have additionally assumed, for simplicity, a constant density profile for this upstream medium, however our results are not particularly sensitive to this choice. E.g. for fiducial FRB parameters and $k=-2$ instead of $k=0$ (where $n_{\rm ext} \propto r^{-k}$) we find that the peak neutrino efficiency $\eta_\nu [t_{\rm max}]$ is only increased by a modest factor of $\sim 5$. Similarly, a positive $k=2$ would decrease the efficiency by a factor of $\sim 4$.}

Although the detection prospects with present neutrino observatories are extremely challenging (to put it lightly), the detection of even a single neutrino from an FRB would be a smoking gun for this model.  
The most promising potentiality is a giant flare from a nearby Galactic magnetar with the fortuitous geometry of the shock-generating plasmoid being directed along our line of sight.   


\acknowledgements
We thank Ali Kheirandish for helpful discussion about IceCube sensitivities to FRBs. We thank Israel Martinez for helpful discussion about HAWC sensitivities to bursts.  BDM acknowledges support from the Simons Foundation (grant 606260). Support for KF was provided by NASA through the NASA Hubble Fellowship grant \#HST-HF2-51407 awarded by the Space Telescope Science Institute, which is operated by the Association of Universities for Research in Astronomy, Inc., for NASA, under contract NAS5-26555. BM is supported by NASA through the NASA Hubble Fellowship grant \#HST-HF2-51412.001-A awarded by the Space Telescope Science Institute, which is operated by the Association of Universities for Research in Astronomy, Inc., for NASA, under contract NAS5-26555.






\end{document}